\documentclass[11pt,reqno,a4,onecolumn,twoside,final,spanish]{article}

\usepackage{amsthm,amsmath,amssymb,latexsym,amsfonts,amscd} 
\usepackage{latexsym}
\usepackage{url}
\usepackage[T1]{fontenc}
\usepackage{ae}
\usepackage{aecompl}
\usepackage[small,sc,hang,bf]{caption} 
\usepackage{graphicx,color,ae,fancyvrb}
\usepackage{pst-all}
\usepackage{subfig}
\usepackage{float} 
\usepackage{verbatim}
\usepackage{enumerate}
\usepackage{array}
\usepackage{longtable}
\usepackage{multicol}
\usepackage{cancel}
\newcommand*\chancery{\fontfamily{pzc}\selectfont}

\newlength\mytemplen
\newsavebox\mytempbox

\makeatletter
\newcommand\mybluebox{%
    \@ifnextchar[
       {\@mybluebox}%
       {\@mybluebox[0pt]}}

\def\@mybluebox[#1]{%
    \@ifnextchar[
       {\@@mybluebox[#1]}%
       {\@@mybluebox[#1][0pt]}}

\def\@@mybluebox[#1][#2]#3{
    \sbox\mytempbox{#3}%
    \mytemplen\ht\mytempbox
    \advance\mytemplen #1\relax
    \ht\mytempbox\mytemplen
    \mytemplen\dp\mytempbox
    \advance\mytemplen #2\relax
    \dp\mytempbox\mytemplen
    \colorbox{myblue}{\hspace{1em}\usebox{\mytempbox}\hspace{1em}}}

\makeatother

\usepackage{natbib}

\usepackage{hyperref}

\usepackage{geometry}
\usepackage{calc}
	
\makeatletter
\renewcommand{\section}{\@startsection{section}{1}{0pt}{-3ex plus -1ex minus 0ex}{2ex plus 0ex}{\bf}}
\makeatother

\makeatletter
\renewcommand{\subsection}{\@startsection{subsection}{1}{0pt}{-2ex plus -1ex minus 0ex}{2ex plus 0ex}{\bf}}
\makeatother

\theoremstyle{definition}

\theoremstyle{remark}

\begin{document}

\renewcommand{\tablename}{Tabla}
\renewcommand{\figurename}{Figura}
\noindent

\begin{flushleft}
\textsl {\chancery  Memorias de la Primera Escuela de Astroestad\'istica: M\'etodos Bayesianos en Cosmolog\'ia}\\
\vspace{-0.1cm}{\chancery  9 al 13 Junio de 2014.  Bogot\'a D.C., Colombia }\\
\textsl {\scriptsize Editor: H\'ector J. Hort\'ua}\\
\href{https://www.dropbox.com/sh/nh0nbydi0lp81ha/AACJNr09cXSEFGPeFK4M3v9Pa}{\tiny {\blue Material suplementario}}
\end{flushleft}



\thispagestyle{plain}\def\@roman#1{\romannumeral #1}



\begin{center}\Large\bfseries Simulaciones Cosmológicas: una herramienta para entender la formación de estructura a gran escala y su conexión con la cosmología Newtoniana \end{center}
\begin{center}\normalsize\bfseries  Cosmological simulations: a tool for understanding the formation of large scale structure and connection with the Newtonian cosmology\end{center}

\begin{center}
\small
\textsc{J.E. García-Farieta\footnotemark[1]}
\textsc{L. Castañeda\footnotemark[2]}
\textsc{J.M. Tejeiro\footnotemark[3]}
\footnotetext[1]{Universidad Nacional de Colombia. E-mail: \url{joegarciafa@unal.edu.co}}
\footnotetext[2]{Universidad Nacional de Colombia - OAN. E-mail: \url{lcastanedac@unal.edu.co}}
\footnotetext[3]{Universidad Nacional de Colombia - OAN. E-mail: \url{jmtejeiros@unal.edu.co}}

\end{center}

\noindent\\[1mm]
{\small
\centerline{\bfseries Resumen}\\
En este documento se estudian las simulaciones cosmológicas como una he\-rramienta para entender la formación de estructura a gran escala del universo.
Se muestra la equivalencia de la cosmología Newtoniana con el gauge de Poisson y se estudia la solución de las ecuaciones de campo de Einstein para este gauge. 
Las simulaciones son imprescindibles para dar cuenta de los resultados del modelo cosmológico en estudio, permitiendo compararlos con datos observacionales. 
Las simulaciones de N-cuerpos se han utilizado desde el estudio sistemas de pocos cuerpos hasta la evolución y formación de estructura no lineal a gran escala como 
filamentos y halos galácticos.\\

{\footnotesize
\textbf{Palabras clave:}
Simulación, estructura a gran escala, cosmología Newtoniana.
\\
\noindent\\[1mm]
{\small
\centerline{\bfseries Abstract}\\

In this work we study the cosmological simulations as a tool to understand the formation of large-scale structure of the universe, 
for this, we show the equivalence of Newtonian cosmology with Poisson gauge and we study the solution of the Einstein's field equations for this gauge.
The simulations are essential for taking account the results of the cosmological model, allowing to compare the results with observational data. 
The N-bodies simulations have been used from systems of few bodies to the evolution and formation of nonlinear large scale structure as filaments and galactic halos.\\

{\footnotesize
}

\newpage
\section{Introducción}

Hoy día la informática se ha convertido en una valiosa herramienta de apoyo a las diferentes disciplinas, especialmente en aquellos campos de investigación donde los modelos teóricos deben proporcionar resultados comparables con experimentos u observaciones. En el caso particular de la cosmología, las simulaciones computacionales han jugado un importante papel en los últimos años. En las últimas décadas se han visto grandes avances en la comprensión de la formación y evolución de estructura cósmica, algunas observaciones recientes de galaxias han permitido cartografiar el universo observable dando cuenta de la estructura galáctica localmente, en tanto los modelos teóricos han permitido explicar cómo estas estructuras surgieron a partir de perturbaciones asociadas al origen del universo. Las enormes escalas de tiempo y distancia involucradas en la formación de estructuras a gran escala no facilitan la comparación directa entre las predicciones del modelo con las observaciones, de allí que la simulación computacional proporcione uno de los mecanismos más importantes en la investigación de la formación de la estructura cósmica a gran escala. Con el fin de realizar simulaciones confiables resulta importante entender el mecanismo y funcionamiento de los diferentes algoritmos de simulación, haciendo énfasis en sus fortalezas y susceptibilidades.\\

Academicamente resulta indudable que los avances en ciencias de la computación, métodos computacionales e implementación de paquetes informáticos han contribuido significativamente a nuestra comprensión de los modelos cosmológicos, el universo, y de procesos astrofísicos. En este documento se parte de las bases teóricas de la relatividad general, se revisan algunos conceptos fundamentales del modelo estándar de la cosmología, y finalmente se muestran algunos resultados de la simulación de una caja cosmológica creada con el código GADGET-2, \cite{Volker1}. 

\section{Ecuaciones de Campo de Einstein Perturbadas en el gauge de Poisson}

La formación de estructura en el universo es uno de los campos de investigación más activos desde el punto de vista teórico y observacional, \cite{CastanedaNotas}; una excelente referencia del tema es el libro \emph{The Large Scale Structure of the Universe} de P.J.E. Peebles, el cual menciona al respecto: ``{\bf \emph{Discussion of how irregularities in the matter distribution behave in an expanding universe is greatly simplified by the fact that a limiting approximation of general relativity, Newtonian mechanics, applies in a small region compared to the Hubble lenght $cH^{-1}$ (and large compared to the Schwarzschild radii of any collapse objects). The rest of the universe can affect the region only through a tidal field. Newtonian approximation is not a model but a limiting case valid no matter what is happening in the distant parts of the universe.}}''\\

La teoría de perturbaciones cosmológicas es una herramienta útil para estudiar las inhomogeneidades primordiales en cosmología, dando cuenta de la formación y evolución de la estructura del Universo, la formación galáctica, el ``clustering'' en la distribución de galaxias y las anisotropías de la radiación de fondo (CMB). El formalismo de la Relatividad General (RG), en la cual se admite la no existencia de un sistema de coordenadas preferencial, corresponde en este sentido a que cualquier cantidad que sea dependiente de las coordenadas implicará que su perturbación sea dependiente del gauge, \cite{Hortua}. La teoría de perturbaciones usa el andamiaje físico-matemático de la RG para calcular las fuerzas gravitacionales ori\-ginadas a partir de las pequeñas perturbaciones, y que al evolucionar dan cuenta de la formación de estrellas, cuásares, galaxias y cúmulos. Estos argumentos sólo son válidos para un universo predominantemente homogéneo. Para un universo suficientemente homogéneo, la teoría de perturbaciones proporciona una buena aproximación a grandes escalas, sin embargo a escalas pequeñas se requieren de técnicas más complejas como las simulaciones de N-cuerpos, \cite{Mukhanov1}.\\

En está sección se obtienen las ecuaciones de campo de Einstein perturbadas bajo el gauge de Poisson, para ello definimos el siguiente elemento de línea

\begin{equation}\label{eq:elementLine}
ds^2=a^2(\tau)[-(1+2\phi)d\tau^2+(1-2\psi)(dx^2+dy^2+dz^2)],
\end{equation}

con los coeficientes métricos

\begin{eqnarray*}\label{eq:MetricComp}
g_{00}&=&-a^2(1+2\phi),\quad\quad g_{0i}=0,\quad\quad g_{ij}=a^2(1-2\psi)\delta_{ij},\\
g^{00}&=&-a^{-2}(1-2\phi),\quad\quad g_{0i}=0,\quad\quad g_{ij}=a^{-2}(1+2\psi)\delta^{ij}.
\end{eqnarray*}

Así la conexión métrica está dada por  $\Gamma^\mu_{\nu\lambda}=\frac{1}{2}g^{\mu\rho}[g_{\mu\rho,\lambda}+g_{\lambda\rho,\nu}-g_{\mu\lambda,\rho}]$. Tras el cálculo se obtienen las componentes del tensor de Ricci expresadas a continuación

\begin{eqnarray*}
R_{00}&=&-3\mathcal{H}'+\nabla^2\phi+3\psi''+3\mathcal{H}\phi'+3\mathcal{H}\psi',\\
R_{0j}&=&2\partial_j(\psi'+\mathcal{H}\phi),\\
R_{ij}&=&[2\mathcal{H}^2-4\phi\mathcal{H}^2-4\psi\mathcal{H}^2-\psi''+(1-2\phi-2\psi)\mathcal{H}'-\mathcal{H}(5\psi'+\phi')+\nabla^2\psi]\delta_{ij}\\&&+\partial_i\partial_j\psi-\partial_i\partial_j\phi.
\end{eqnarray*}

El escalar de Ricci $R$ se calculó como la traza del tensor Ricci, con lo que resulta

\begin{equation*}
R=g^{\mu\nu}R_{\mu\nu}=2a^{-2}(3\mathcal{H}'+3\mathcal{H}^2-3\psi''-\nabla^2\phi+2\nabla^2\phi-3\mathcal{H}\phi'-9\mathcal{H}\psi'-6\mathcal{H}'\phi-6\mathcal{H}^2\phi).
\end{equation*}

Basados en los resultados anteriores es sencillo encontrar las componentes del tensor de Einstein $G_{\mu\nu}=R_{\mu\nu}-\frac{1}{2}g_{\mu\nu}R,$ siendo sus elementos

\begin{eqnarray*}
G_{00}&=&2\nabla^2\psi-6\mathcal{H}\psi'+3\mathcal{H}^2,\\
G_{0j}&=&2\partial_j(\psi'+\mathcal{H}\phi),\\
G_{ij}&=&[2\mathcal{H}^2-4\phi\mathcal{H}^2-4\psi\mathcal{H}^2-\psi''+(1-2\phi-2\psi)\mathcal{H}'-\mathcal{H}(5\psi'+\phi')+\nabla^2\psi]\delta_{ij}\\
&&+\partial_i\partial_j\psi-\partial_i\partial_j\phi-\frac{1}{2}[-a^{2}(1-2\psi)]\delta_{ij}]2a^{-2}(3\mathcal{H}'+3\mathcal{H}^2-3\psi''-\nabla^2\phi+2\nabla^2\phi\\
&&-3\mathcal{H}\phi'-9\mathcal{H}\psi'-6\mathcal{H}'\phi-6\mathcal{H}^2\phi)(-\mathcal{H}^2-2\mathcal{H}'+2\mathcal{H}^2\phi+2\mathcal{H}^2\psi+6\mathcal{H}'\phi\\
&&+4\mathcal{H}'\psi+4\mathcal{H}\psi'+4\mathcal{H}\phi'-\nabla^2\psi+\nabla^2\phi+2\psi'')\delta_{ij}+\partial_i\partial_j(\psi-\phi).
\end{eqnarray*}

De otra parte las componentes del tensor momentum-energía est\'an dadas por ($T^\mu_\nu=T^\mu_{(0)\nu}+\delta T^\mu_\nu$)

\begin{eqnarray*}
T_{00}&=&a^2(\rho+\delta\rho+2\rho\phi),\\
T_{0j}&=&-a^2\delta T^0_j=0,\\
T_{ij}&=&a^2(P+\delta P-2P\phi).
\end{eqnarray*}

Es de resaltar que la componente $0j$ del tensor de momentum-energía es igual a cero dado que no hay disipación de energía a primer orden. Se procede a estudiar las componentes de campo perturbadas $\delta G_{\mu\nu}=8\pi G\delta T_{\mu\nu}$,
con $\mu=0,~\nu=0$

\begin{eqnarray}\label{eq:FielPertu1}
\delta G_{00}&=&8\pi G\delta T_{00}\nonumber\\,
\Rightarrow\nabla^2\psi&=&3\mathcal{H}\psi'+4\pi Ga^2\delta\rho+8\pi Ga^2\rho\phi.
\end{eqnarray}

Con $\mu=0,~\nu=j$

\begin{eqnarray}\label{eq:FielPertu2}
\delta G_{0j}&=&8\pi G\delta T_{0j}\nonumber\\,
\Rightarrow2\partial_j(\psi'+\mathcal{H}\phi)&=&0\nonumber\\,
\Rightarrow\psi'&=&-\mathcal{H}\phi.
\end{eqnarray}

Reemplazando la ecuación (\ref{eq:FielPertu2}) en (\ref{eq:FielPertu1}) se obtiene

\begin{equation}\label{eq:PrePoisson0}
\nabla^2\psi=-3\mathcal{H}^2\phi+4\pi Ga^2\delta\rho+8\pi Ga^2\rho\phi.
\end{equation}

Ahora, teniendo en cuenta la ecuación de Friedmann correspondiente a orden cero en el tiempo y reemplazando éste resultado en la ecuación (\ref{eq:PrePoisson0}) se obtiene

\begin{eqnarray}\label{eq:PrePoisson1}
\nabla^2\psi=4\pi Ga^2\delta\rho.
\end{eqnarray}

Como es de notar la ecuación (\ref{eq:PrePoisson1}) es la ecuación de campo de Einstein tipo Poisson, la cual es congruente con los resultados de la cosmología Newtoniana salvo un factor de escala $a$.

\section{Simulaciones Cosmológicas: El Código GADGET-2 y la física asociada.}\label{sec:Simulacion}

Actualmente el estudio del universo a través de diversos modelos cosmológicos demanda la creación de software robustos y de fácil aplicación que permitan dar cuenta de sus resultados, pudiendo ser comparados con datos observacionales; esto conlleva inmediatamente a resolver numéricamente los sistemas de ecuaciones adecuados para la dinámica de N-cuerpos que interactúan gravitacionalmente. En particular una simu\-lación computacional de N-cuerpos es un software informático cuyo fin es recrear un modelo abstracto para las partículas que componen un sistema físico; los resultados de la simulación proporcionan información acerca de la  dinámica de las partículas junto con otras propiedades de interés del sistema. Las simulaciones computacionales de N-cuerpos son una herramienta ampliamente utilizada para varios fines, tanto en astrofísica como en cosmología, desde el estudio sistemas que comprenden ``pocos'' cuerpos como sistemas solares, hasta la evolución y formación de estructura no lineal a gran escala del universo como filamentos y halos galácticos. Las simulaciones computacionales constituyen una poderosa e importante herramienta para comprender el cosmos; una simulación detallada puede proporcionar un medio para comprender los procesos que se producen en escalas de tiempo y distancias cosmológicas, 

Las actuales simulaciones cosmológicas tienen precedentes que se remontan a varias décadas atrás. La primera simulación de N-cuerpos registrada se realizó en 1941 por Erik Holmberg, \cite{Bertschinger}. Los resultados fueron presentados en el artículo ``\emph{On the Clustering Tendencies among the Nebulae. II. a Study of Encounters Between Laboratory Models of Stellar Systems by a New Integration Procedure}''. La simulación se realizó utilizando una computadora analógica óptica en donde se simu\-lan dos galaxias elípticas interactuando en un plano (ver figura \ref{fig:Holmberg}). Holmberg simuló el sistema de partículas gravitantes como una distribución de 37 bombillas móviles, cuyas posiciones se registraban en una hoja de papel cuadriculado, \cite{Holmberg}.  Dado que la intensidad de luz de una fuente puntual disminuye como $1/r^2$, ésta se puede relacionar directamente con la aceleración de la gravedad en cada punto. El cálculo de la fuerza gravitacional sobre una partícula, debido a la acción de las demás partículas, se reduce a una medida de la intensidad total de la luz, ésta última se determinaba experimentalmente utilizando fotoceldas y galvanómetros.\\

\begin{figure}[h]
    \centering
    \includegraphics[width=0.6\textwidth]{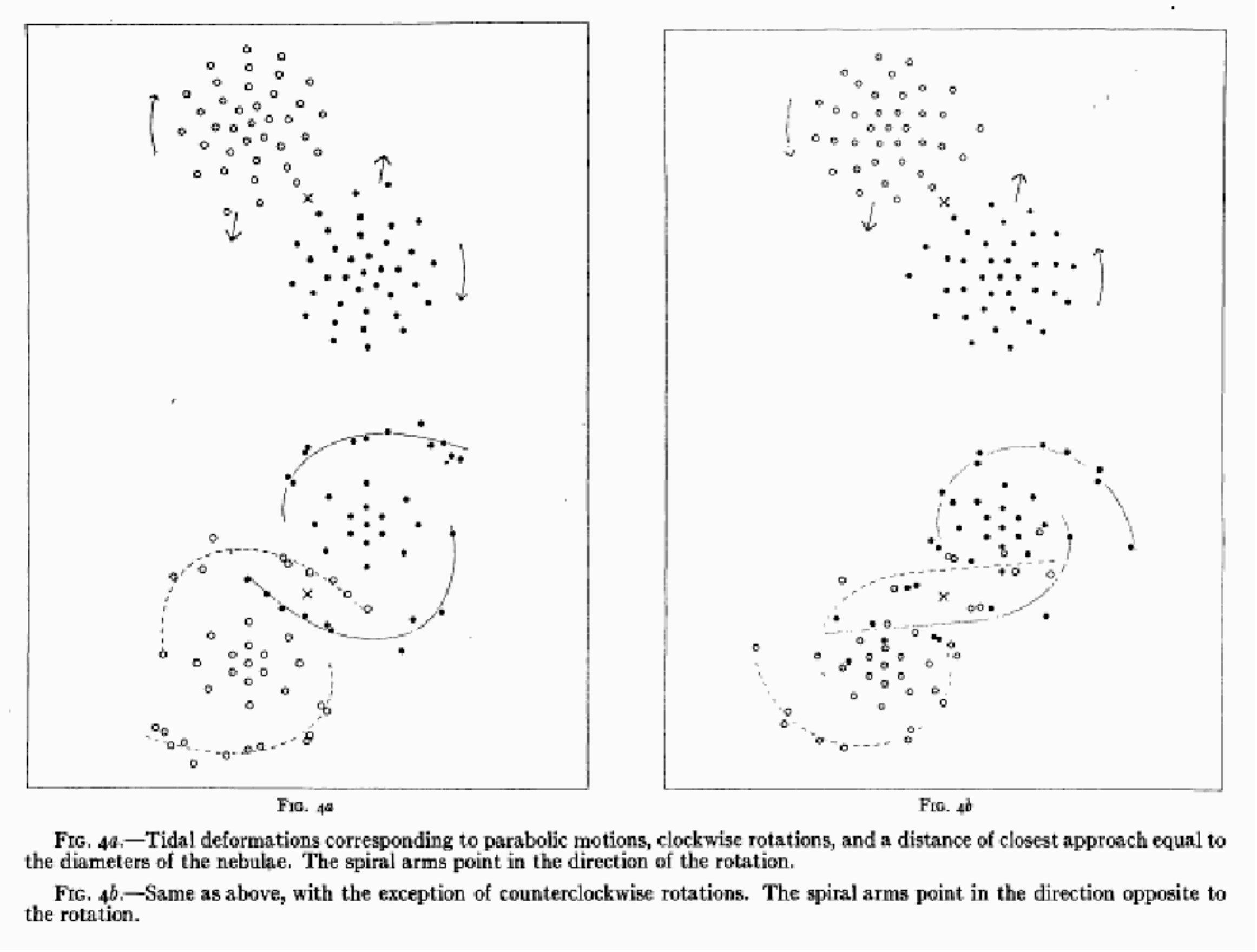}
    \caption{Resultados de las primeras simulaciones cosmol\'ogicas, \cite{Holmberg}.}
    \label{fig:Holmberg}
\end{figure}

En los últimos años se ha visto un crecimiento significativo, no sólo en el tamaño de las simulaciones cosmológicas, sino también en la optimización de los algoritmos físicos implementados. La dinámica de gases ha llegado al punto en que es posible comparar los resultados computacionales con las observaciones. La evolución en el número de partículas en simulaciones cosmológicas de N-cuerpos en función del tiempo (en años) se aprecia en la figura \ref{fig:MooreLaw}, en donde se resaltan algunas de las simulaciones que han tenido mayor impacto. 

\begin{figure}[h]
    \centering
    \includegraphics[width=0.7\textwidth]{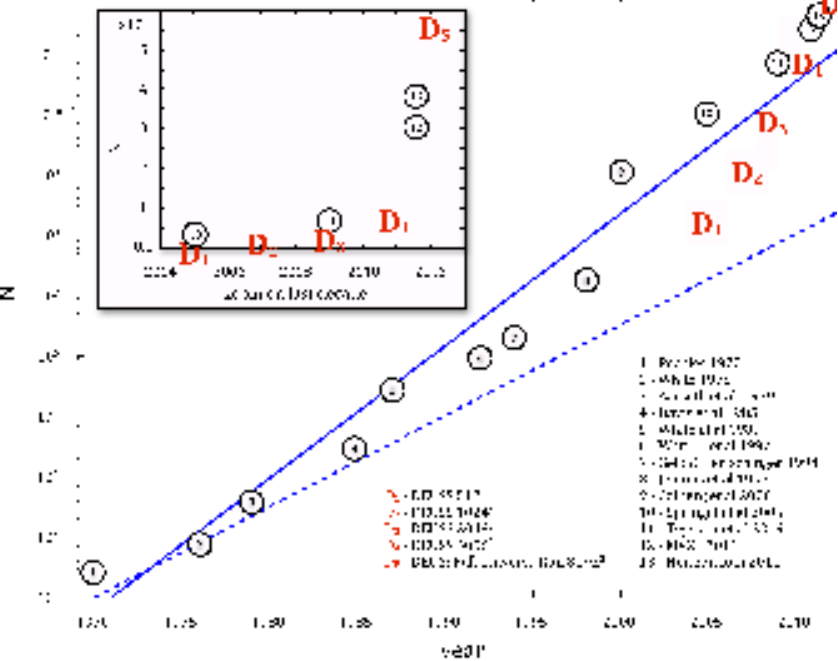}
    \caption{Analog\'ia con la ley de Moore en la que se representa la potencia de computo. La l\'inea azul s\'olida es la evolución media del tama\~no de simulaci\'on, y la l\'inea azul punteada es la Ley de Moore equivalente a simulaciones cosmol\'ogicas, en la que se demuestra un aumento en un factor de 2 en el tama\~no de la simulaci\'on cada 18 meses.}
    \label{fig:MooreLaw}
\end{figure}

\subsection{Estructura y funcionamiento del código}

El código GADGET (por sus siglas en inglés {\bf GA}laxies with {\bf D}ark matter and {\bf G}as int{\bf E}rac{\bf T}), es un software libre, de código abierto, con licencia GNU GPL, inicialmente escrito por Volker Springel, \cite{Volker1}, que permite realizar simulaciones cosmológicas N-body/SPH (Smoothed-particle hydrodynamics) sin colisiones entre partículas y con interacción de gases. La relevancia del código GADGET ra\-dica en que está elaborado con un algoritmo de árbol jerárquico mediante el cual calcula las fuerzas gravitacionales (opcionalmente utiliza un esquema de grilla de partículas para calcular fuerzas gravitacionales de largo alcance, \cite{Volker2}), y representa fluidos por medio de hidrodinámica de partículas suavizadas (SPH). La simulación de estrellas y materia oscura para un sistema autogravitante se modela hidrodinámicamente haciendo uso de la ecuación de Boltzmann no colisional, ver ecuación (\ref{eq:Boltzmann1}), que describe la evolución de un gas sujeto a fuerzas externas, \cite{Barnes}. Las ecuaciones de movimiento según la gravitación Newtoniana, se pueden escribir para este sistema como

\begin{eqnarray}\label{eq:NewtonBoltz1}
\frac{d\vec{r}_i}{dt}=\vec{v}_i,\\
\frac{d\vec{v}_i}{dt}=\sum^N_{j\neq i}Gm_j\frac{\vec{x}_j-\vec{x}_i}{|\vec{x}_j-\vec{x}_i|^3},
\end{eqnarray}

donde se han considerado $N$ masas puntuales, $\vec{x}_i$ corresponde a la posición de la i-ésima partícula, $\vec{v}_i$ a la velocidad de la i-ésima partícula, y $m$ a la masa correspondiente de la partícula en cuestión. Sin embargo, resulta más  conveniente adoptar un tratamiento continuo del sistema, dado que si se tienen en cuenta los valores típicos para un cúmulo globular o para una galaxia promedio, es claro que las ecuaciones de Newton son bastante generales, \cite{Barnes}. En la descripción continua no es necesario especificar las masas, posiciones y velocidades de cada una de las $N$ partículas; en cambio, se define una función de distribución de masas en un espacio de fase $6N$ dimensional. La masa en un punto en un tiempo $t$ se puede definir en términos de la función de distribución o de la densidad del espacio fase como $f(\vec{x},\vec{v},t)d\vec{x}d\vec{v}=\text{masa}$ en un elemento de $d\vec{x}d\vec{v}$.\\

Por otra parte, podemos distinguir entre las estrellas y la materia oscura, escribiendo la función de distribución como $f=f_s+f_d$. Donde $f_s$ es la función de distribución asociada a las estrellas y $f_d$ es la función de distribución asociada a la materia oscura. Vale aclarar que todos los posibles valores asociados a la masa quedan confinados en el espacio de fase. Con el fin de encontrar la ecuación dinámica de la función de distribución, suponemos que el flujo de materia a través del espacio de fase $6N$ dimensional se rige por un campo vectorial también de 6 dimensiones, de tal manera que $(\dot\vec{x},\dot\vec{v})=(\vec{v},-\nabla\phi)$, donde $\phi(\vec{x},t)$ es el potencial gravitacional. Teniendo en cuenta la conservación de la masa se tiene que

\begin{equation}\label{eq:Boltzmann1}
\frac{\partial f}{\partial t}+\frac{\vec{p}}{m}\cdot\vec{\nabla}f+\frac{\vec{F}}{m}\cdot\vec{\nabla}_vf=\left(\frac{\partial f}{\partial t}\right)_{col}=0.
\end{equation}

Para fuerzas conservativas tal que $\vec{F}=-m\vec{\nabla}\phi$ (con $m$ la masa de la partícula), la ecuación anterior se escribe como

\begin{equation}\label{eq:Boltzmann2}
\frac{\partial f}{\partial t}+\vec{v}\cdot\vec{\nabla}f-\vec{\nabla}\phi\cdot\vec{\nabla}_vf=0.
\end{equation}

Esta última ecuación es la ecuación de Boltzmann no colisional, también llamada ecuación Vlasov  y es un caso particular del teorema de Liouville.  Esencialmente, la ecuación (\ref{eq:Boltzmann2}) establece que el flujo de partículas a través de una región del espacio de fase es incompresible, o que la densidad del espacio de fase alrededor de un punto que representa una estrella cualesquiera, permanece constante. Además se ha de cumplir que el potencial gravitacional satisfaga la ecuación de Poisson

\begin{equation}\label{eq:Poisson}
\nabla^2\phi(\vec{x},t)=4\pi G\int_Sf(\vec{x},\vec{v},t)d^3v,
\end{equation}

donde $S$ representa todo el espacio de velocidades, y donde $f$ igualmente se define como $f(\vec{r},\vec{v},t)d\vec{v}d\vec{x}$ que viene dada por la masa total de las partículas que se encuentran en un volumen $d^3x$ con radio vector $\vec{x}$ y velocidad también ubicada en un pequeño cubo de volumen $d^3v$, con radio vector $\vec{v}$.\\

Las ecuación de Poisson junto con la ecuaci\'on (\ref{eq:Boltzmann1}) conforman la totalidad de las ecuaciones necesarias para describir la dinámica de un gas no colisional autogravitante en el caso de un universo Newtoniano. Sin embargo, dada la complejidad de resolver directamente las ecuaciones acopladas se recurre al método de N-cuerpos, en el que $f$ pasa a ser una serie de partículas representativas, dadas por deltas de Dirac, \cite{Volker1}. En este esquema se introduce el suavizamiento gravitacional $\epsilon$ que permite evaluar los saltos abruptos de la fuerza requiriendo un intervalo de avance de tiempo sumamente pequeño para seguir correctamente la dinámica de dicha situación. Para el caso de un sistema de dos partículas aisladas que interactúan mediante un potencial Newtoniano, se tiene que la fuerza entre ellas esta descrita por la ecuaci\'on (\ref{eq:GadgeFuerza1}), con una divergencia cuando la distancia entre dos partículas es cero, lo que produce aceleraciones muy altas; con ello, sí $\vec{x}_i\rightarrow\vec{x}_j$, metodológicamente se requeriría de un paso infinitesimalmente cada vez mas pequeño

\begin{equation}\label{eq:GadgeFuerza1}
\vec{F}_i=-\sum_{j\neq i}G\frac{m_im_j(\vec{x}_i-\vec{x}_j)}{|\vec{x}_i-\vec{x}_j|^3}.
\end{equation}

Este problema se soluciona agregando el t\'ermino $\epsilon$ como se observa en la ecuación (\ref{eq:GadgeFuerza2}), siendo $\epsilon$ el \emph{softening lenght},  \cite{Bodenheimer}. En este caso, la fuerza puede escribirse como

\begin{equation}\label{eq:GadgeFuerza2}
\vec{F}_i=-\sum_{j\neq i}G\frac{m_im_j(\vec{x}_i-\vec{x}_j)}{(|\vec{x}_i-\vec{x}_j|^2+\epsilon^2)^{3/2}}.
\end{equation}

Físicamente se podría interpretar esta longitud de suavizado $\epsilon$ como la distancia entre los centros de dos partículas que están ``unidas''. De otra parte existen diversos algoritmos incorporados en el código GADGET-2, que resuelven la fuerza gravitacional de manera mas aproximada y con la mayor eficiencia posible. En el caso descrito en un comienzo, el potencial puede escribirse de la manera usual de acuerdo con la ecuación (\ref{eq:Poisson})  e introduciendo el suavizado $\epsilon$ por lo tanto el potencial queda descrito por

\begin{equation}\label{eq:Poisson1}
\phi(\vec{x},t)=-G\int_s\int_s\frac{f(\vec{x}',\vec{v}',t)d^3v'd^3x'}{|\epsilon^2+\vec{x}-\vec{x}'|}.
\end{equation}

La evolución de la dinámica del sistema de N-cuerpos, con partículas autogravitantes y teniendo en cuenta el factor de escala $a(t)$, se describe mediante el Hamiltoniano mostrado en la ecuaci\'on (\ref{eq:Gadge1Hamilton}),  \cite{Volker1},  donde $\vec{p}_i$,  $\vec{x}_i$, son vectores des\-critos en coordenadas comóviles y el correspondiente momentum canónico esta dado por $\vec{p}_i=a(t)^2m_i\dot{\vec{x}}$. La dependencia temporal del Hamiltoniano está en la evolución del factor de escala, el cual se describe mediante el modelo de Friedmann-Lema\^itre. En el caso de un sistema cuya dinámica se describa en un espacio Newtoniano, i.e. $a(t)=1$, el Hamiltoniano correspondiente es independiente del tiempo, estando el potencial descrito por la forma usual $G/|\vec{x}|$, y la fuerza por la ecuación (\ref{eq:GadgeFuerza1}). En  casos para los cuales $a(t)\neq1$, la evolución temporal se determina directamente de la ecuación de Friedmann, eligiéndose el modelo deseado, que corresponde en este caso a uno de los universos del modelo de Friedmann-Lema\^itre, que son simplemente los tipos de universos descritos por la ecuación de Friedmann.

\begin{equation}\label{eq:Gadge1Hamilton}
H=\sum_i\frac{p_i^2}{2m_ia(t)^2}+\frac{1}{2}\sum_{ij}\frac{m_im_j\phi(x_i-x_j)}{a(t)}
\end{equation}

Consideremos ahora el caso de una caja periódica de lado $L$; el potencial escalar $\phi$ viene escrito como 

\begin{equation}\label{eq:}
\nabla^2\phi(\vec{x})=4\pi G\left[-\frac{1}{L^3}+\sum_{\vec{n}}W(|\vec{x}-\vec{n}L, 2.8\epsilon|)\right],
\end{equation}

siendo $\vec{n}=(n_1, n_2, n_3)$ un vector extendido sobre todos las triplas de números enteros, y $W$ una función continua que se hace cero a partir de un punto determinado, y que depende de $\epsilon$. El término $W$ se conoce usualmente como el kernel de una función, y representa una forma de medir qué tan ``lejos'' está una función de ser uno a uno, con ello establece una relación de equivalencia en el dominio de dicha función. El t\'ermino de discretización  $2.8\epsilon$ corresponde al esquema del kernel de discretización de Monaghan y Lattanzio (1985) usado en SPH,  \cite{Volker2}. El ``potencial peculiar'' estará definido como $\phi(\vec{x})=\sum_im_i\varphi(\vec{x}-\vec{x}_i)$, que satisface la ecuación de Poisson correspondiente a un campo de fluctuaciones dado por

\begin{equation}\label{eq:Posifluctua}
\nabla^2\phi(\vec{x})=4\pi G\left[\rho(\vec{x})-\bar\rho(\vec{x})\right].
\end{equation}

Esto concluye la descripción de como se modelan estrellas y materia oscura. Res\-pecto al medio interestelar o al medio intergaláctico, estos se pueden tratar como  un fluido ideal, \cite{Volker1}. El código GADGET-2 utiliza SPH (hidrodinámica de partículas lisas). Este método utiliza un conjunto de partículas representativas para describir el estado del fluido. Las partículas con coordenadas $\vec{x}_i$ , velocidades $\vec{v}_i$ y masas $\vec{m}_i$ son modeladas como elementos de un fluido que representan un gas en un modo Lagrangiano, \cite{Volker2}. El estado termodinámico de cada elemento del fluido puede ser definido en términos de su energía térmica por unidad de masa o en términos de la entropía por unidad de masa. Se sabe que el estado del fluido queda completamente descrito por las ecuaciones de la hidrodinámica, \cite{Longair}, en este caso, las ecuaciones de Euler que describen el movimiento de un fluido compresible no viscoso, en conjunto con la ecuación de Poisson

\begin{eqnarray}\label{eq:BasicHidrody}
\frac{\partial\rho}{\partial t}+\vec{\nabla}\cdot(\rho\vec{v})&=&0,\\
\frac{\partial\vec{v}}{\partial t}+(\vec{v}\cdot\nabla)\vec{v}&=&-\frac{1}{\rho}\vec{\nabla}(p)-\vec{\nabla}\phi,\\
\vec{\nabla}^2\phi=4\pi G\rho.
\end{eqnarray}

En este esquema, la velocidad del fluido $\vec{v}$ es un campo vectorial que depende de la posición y del tiempo, lo cual permite una visión ``global'' del fluido. Por otro lado, est\'a la conocida representación de Lagrange, en la que se elige un punto del campo vectorial obtenido por el esquema de Euler en algún tiempo
$t=t_0$, y evoluciona en el tiempo, lo que permite estudiar la dinámica de puntos individuales del fluido. Para pasar a esta representación expresamos el operador derivada total como $\frac{d}{dt}=\frac{\partial}{\partial t}+\vec{v}\cdot\vec{\nabla}$, transformándose las ecuaciones anteriores en

\begin{eqnarray}\label{eq:EulerFluid}
\frac{d\rho}{dt}&=&-\rho\vec{\nabla}\cdot\vec{v},\\
\frac{d\vec{v}}{dt}&=&-\frac{1}{\rho}\vec{\nabla}p-\vec{\nabla}\phi,\\
\vec{\nabla}^2\phi&=&4\pi G\rho.
\end{eqnarray}

Además, la ecuación de conservación de energía se expresa a partir de la primera ley de la termodinámica como

\begin{equation}\label{eq:LagranFluid}
\frac{du}{dt}+\frac{P}{\rho}\vec{\nabla}\cdot\vec{v}=0,
\end{equation}

en donde $u$ es la energía por unidad de masa, y no se han considerado términos de fuentes, ya que el lado derecho es cero. En lo que concierne a este documento basta considerar que la energía se conserva. Para un gas ideal se cumple también que

\begin{equation}\label{eq:IdealGas}
P=(\gamma-1)\rho u,
\end{equation}
 
en tal caso para gases monoatómicos se considera $\gamma=5/3$. Para modelar este fluido se usan nuevamente partículas representativas para describir $\rho$, pero ahora implementando lo que se denomina Smoothed Particle Hydrodynamics. La idea básica es la siguiente: se define la interpolación integral de una función arbitraria $A(\vec{x})$ como

\begin{equation}\label{eq:Kernel0}
A_I(\vec{x})=\int_S A(\vec{x}')W(\vec{x}-\vec{x}')d^3x,
\end{equation}

donde a $W$ es el kernel de interpolación para la función $A(\vec{x}')$, el cual debe satisfacer que $\lim_{h\rightarrow0}W(\vec{x}-\vec{x}',h)=\delta(\vec{x}-\vec{x}')$. Vale recordar que en el caso de partículas cuya densidad de masa este dada por funciones deltas de Dirac, la densidad en todo el espacio estará dada por

\begin{equation}\label{eq:Kernel1}
\rho(\vec{x})=\sum^N_{j=1}m_jW(\vec{x}-\vec{x}',h).
\end{equation}

De acuerdo con \cite{Volker2}, un buen núcleo de interpolación es el siguiente

\begin{equation}\label{eq:Kernel2}
W(\vec{x}-\vec{x}',h)=\frac{1}{h^3\pi^{3/2}}e^{-|\vec{x}-\vec{x}'|^2/h^2}.
\end{equation}

El resultado es el suavizamiento de la densidad, lo que permite por ejemplo en el caso de un kernel gaussiano como el anterior, describir una densidad continua en todo el espacio. Una vez expresada la densidad, para el cálculo numérico, la interpolación integral de una cierta cantidad ya sea la presión u otra variable, se hace mediante una aproximación de la integral expresada por la ecuación (\ref{eq:Kernel0}), \cite{Volker2}

\begin{equation}\label{eq:AproIntegralA}
A_s(\vec{x})=\sum_jm_j\frac{A_j}{\rho_j}W(\vec{x}-\vec{x}',h),
\end{equation}

donde $A_j=A(\vec{x}_j)$ y $\rho_j=\rho(\vec{x}_j)$. Sólo basta agregar que el kernel usado en GADGET-2 perteneciente a la clase de kernels de soporte compacto, \cite{Volker2}.  Específicamente, se utiliza el kernel dado en la ecuación (\ref{eq:Kernel3}); una revisión detallada se presenta en la referencias ya comentadas, en particular en \cite{Volker2}.

\begin{equation}\label{eq:Kernel3}
W(r,h)=\frac{8}{\pi h^3}\begin{cases}
1-6\left(\frac{|\vec{x}|}{h}\right)^2+6\left(\frac{|\vec{x}|}{h}\right)^3& 0\leq\frac{|\vec{x}|}{h}<\frac{1}{2}\\
2\left(1-\frac{|\vec{x}|}{h}\right) & \frac{1}{2}\leq\frac{|\vec{x}|}{h}<1\\
0 & \frac{|\vec{x}|}{h}>1\end{cases}
\end{equation}

\section{Resultados: simulación de una caja cosmológica}
Los modelos de formación de estructura están estrechamente relacionados en cosmología con un problema de valor inicial. Dadas las condiciones iniciales correspondientes a un fondo determinado por el modelo cosmológico, con la composición especificada de materia, radiación, campos, constante cosmológica y fluctuaciones primordiales en la materia, la radiación, y la geometría del espacio-tiempo.  El objetivo de la simulación es visualizar la formación y evolución de estructuras desde una época en la que el universo era regularmente homogéneo, hasta nuestros días.\\

La simulación ejecutada con este código corresponde a un universo espacialmente plano en expansión, con los siguientes parámetros: 16777215 partículas, densidad total de materia en $z=0~$ $\Omega_0=0.227$, constante cosmológica en $z=0~$ $\Omega_\Lambda=0.728$, densidad de materia bariónica en $z=0~$ $\Omega_m=0.045$. La simulación tiene condiciones de contorno de periódicas dado que el sistema es de gran tamaño. Los resultados obtenidos son bastante similares a simulaciones cosmológicas como las citadas en \cite{Volker1}, realizadas por el \emph{Virgo Consortium}, en particular la \emph{Millenium Simulation}. En los resultados obtenidos (ver figuras \ref{fig:Simul2} y \ref{fig:Simul3}), se aprecia cómo a partir de una distribución cuasi-homogénea de partículas, la cual se puede ver directamente hacia redshift cercano a $z=63$, las partículas se dispersan formando diversos cúmulos aproximadamente en un redshift $0.5$, finalmente se forman varias ramificaciones entre estos cúmulos cerca de $z=0$ hasta que visualmente están interconectados unos con otros.\\

\begin{figure}[htb]
    \centering 
    \includegraphics[width=0.3\textwidth]{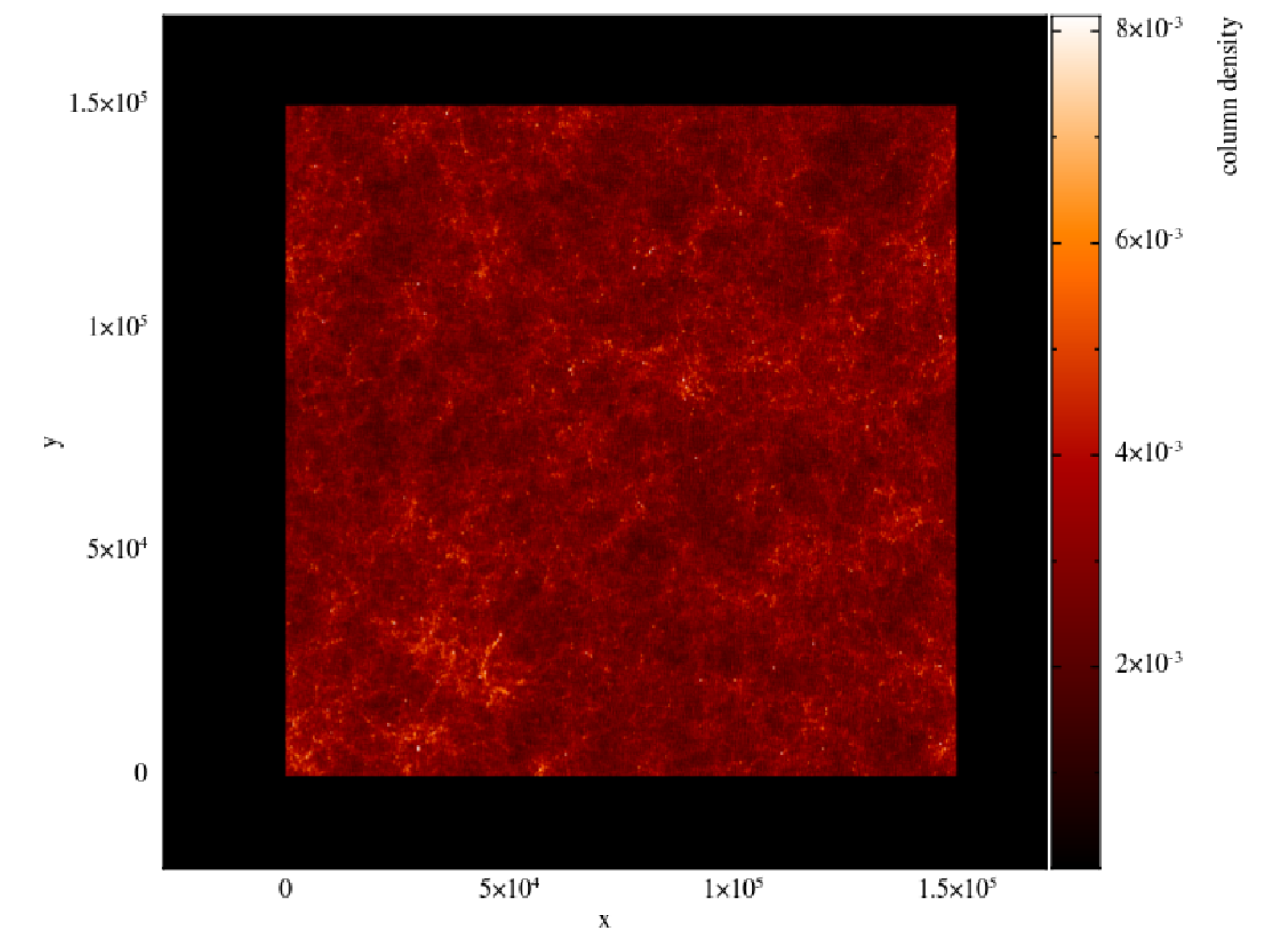}
    \includegraphics[width=0.3\textwidth]{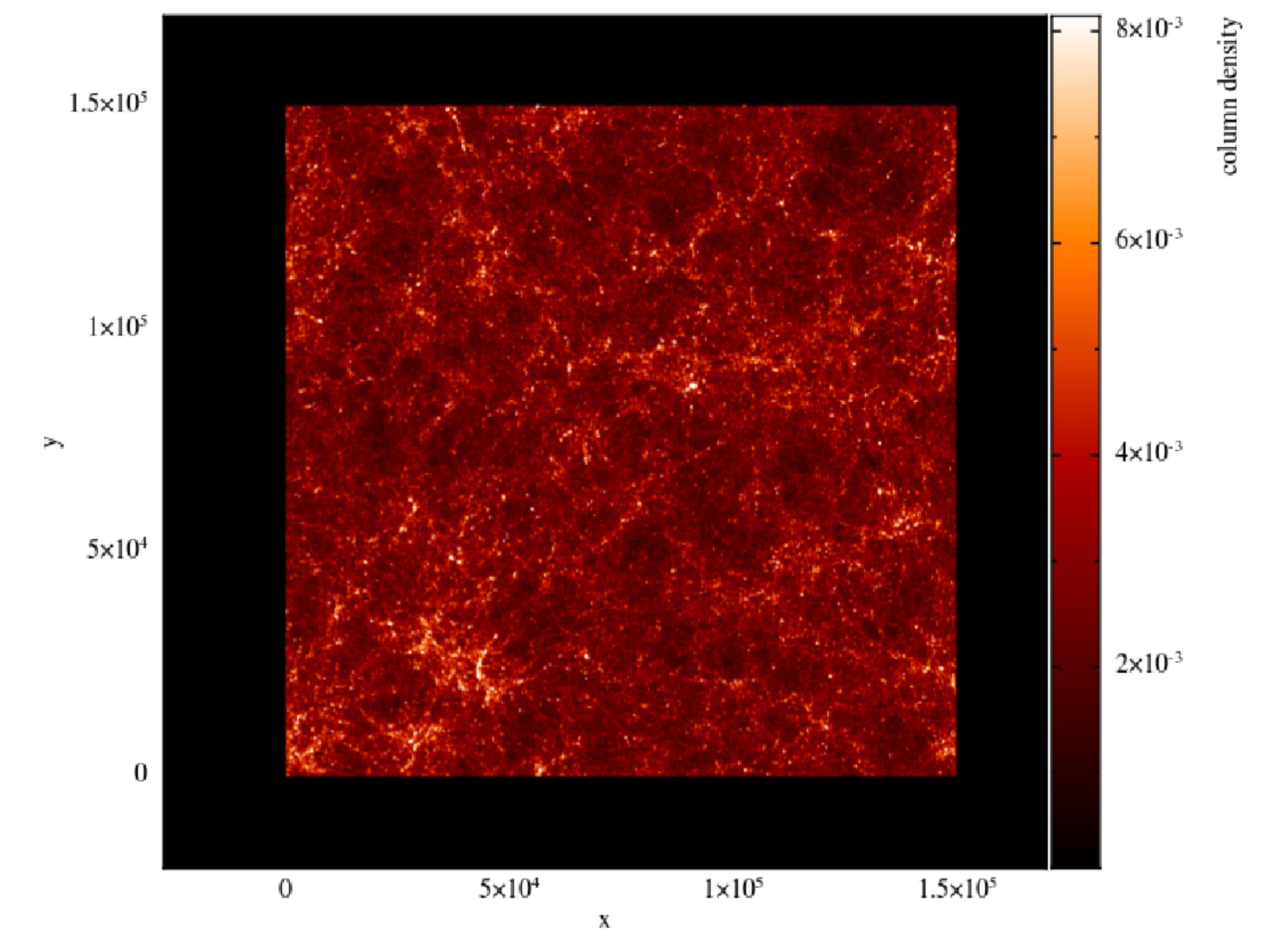}
    \includegraphics[width=0.3\textwidth]{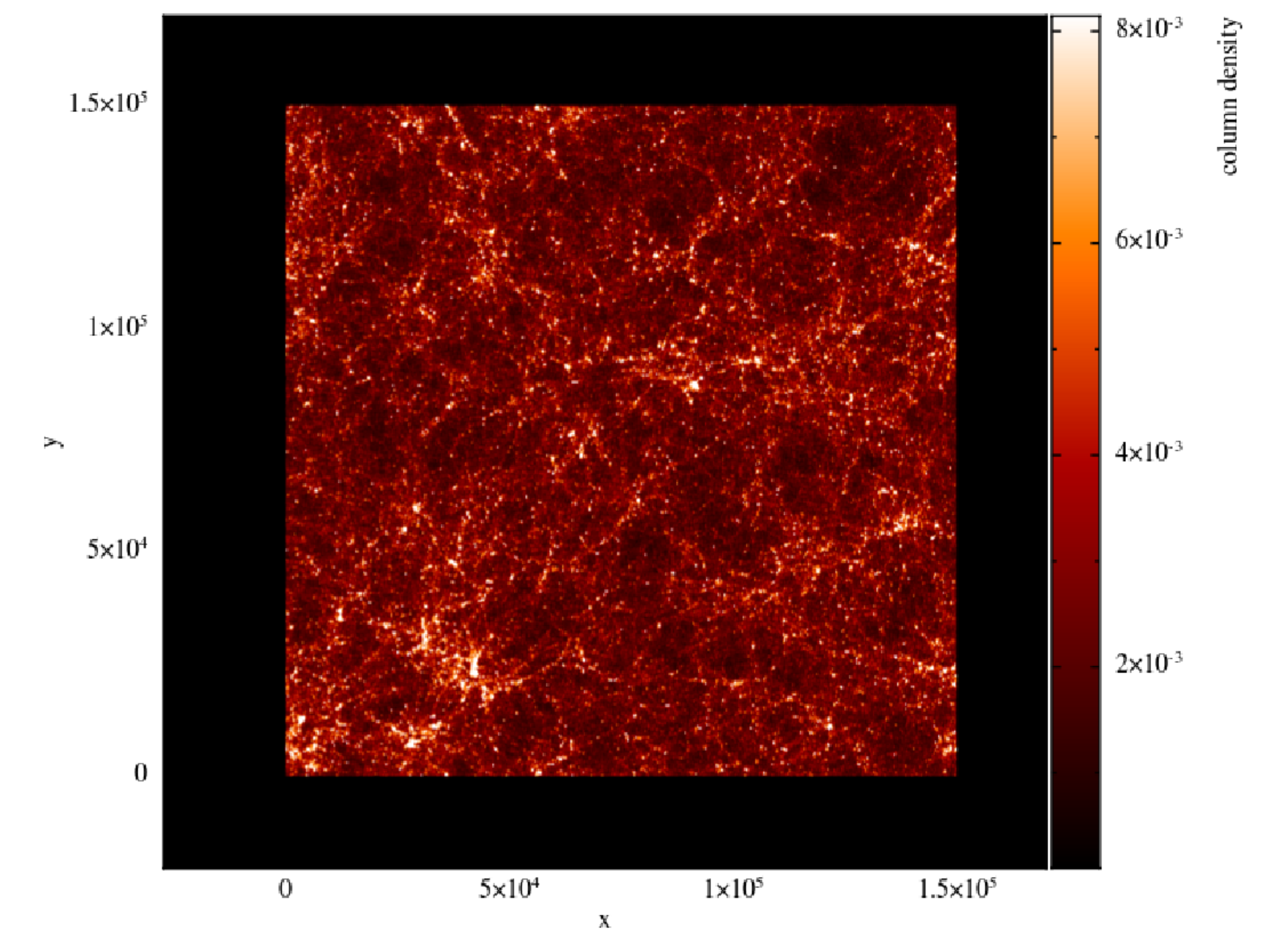}
    \includegraphics[width=0.3\textwidth]{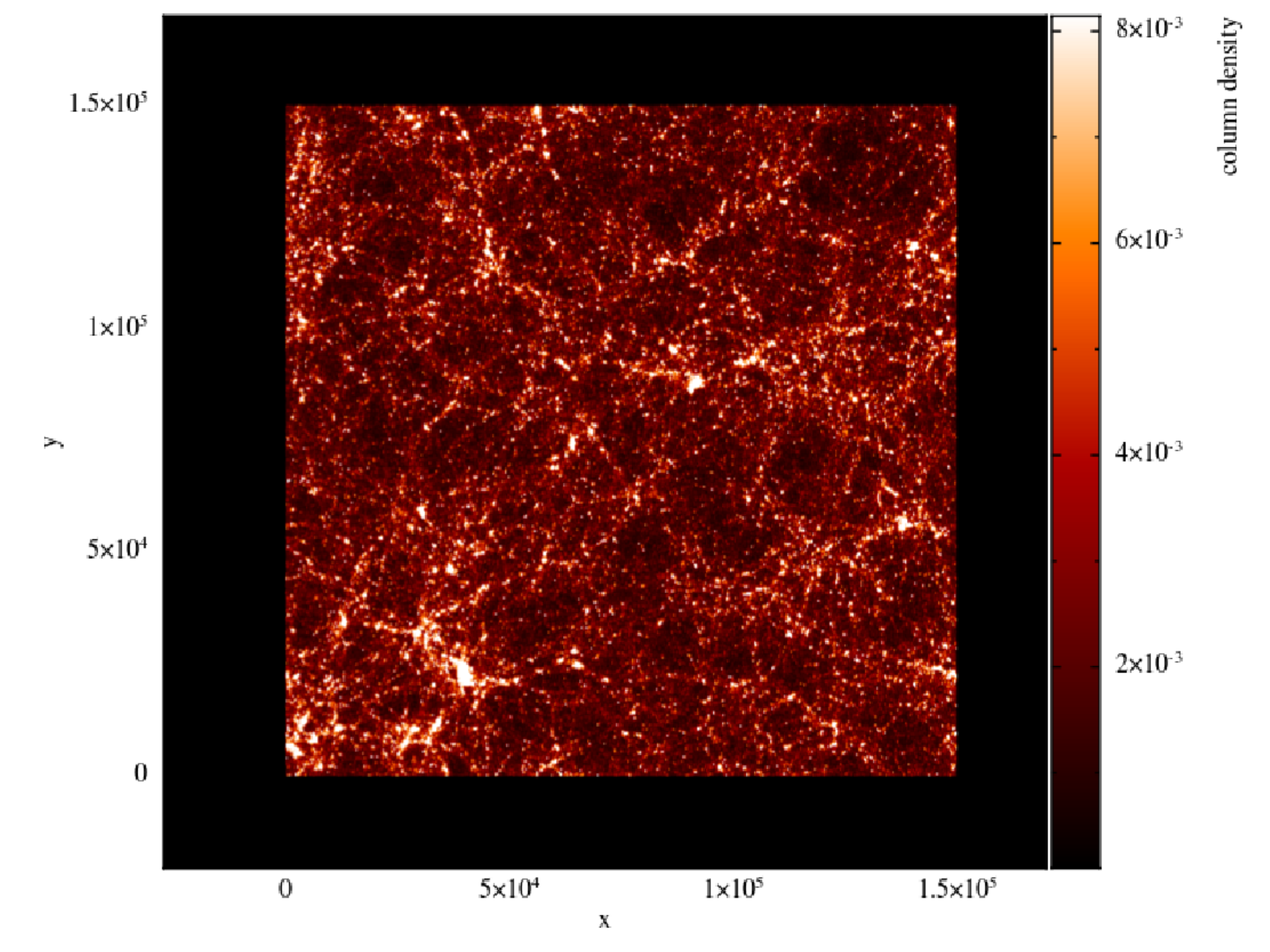}
    \includegraphics[width=0.3\textwidth]{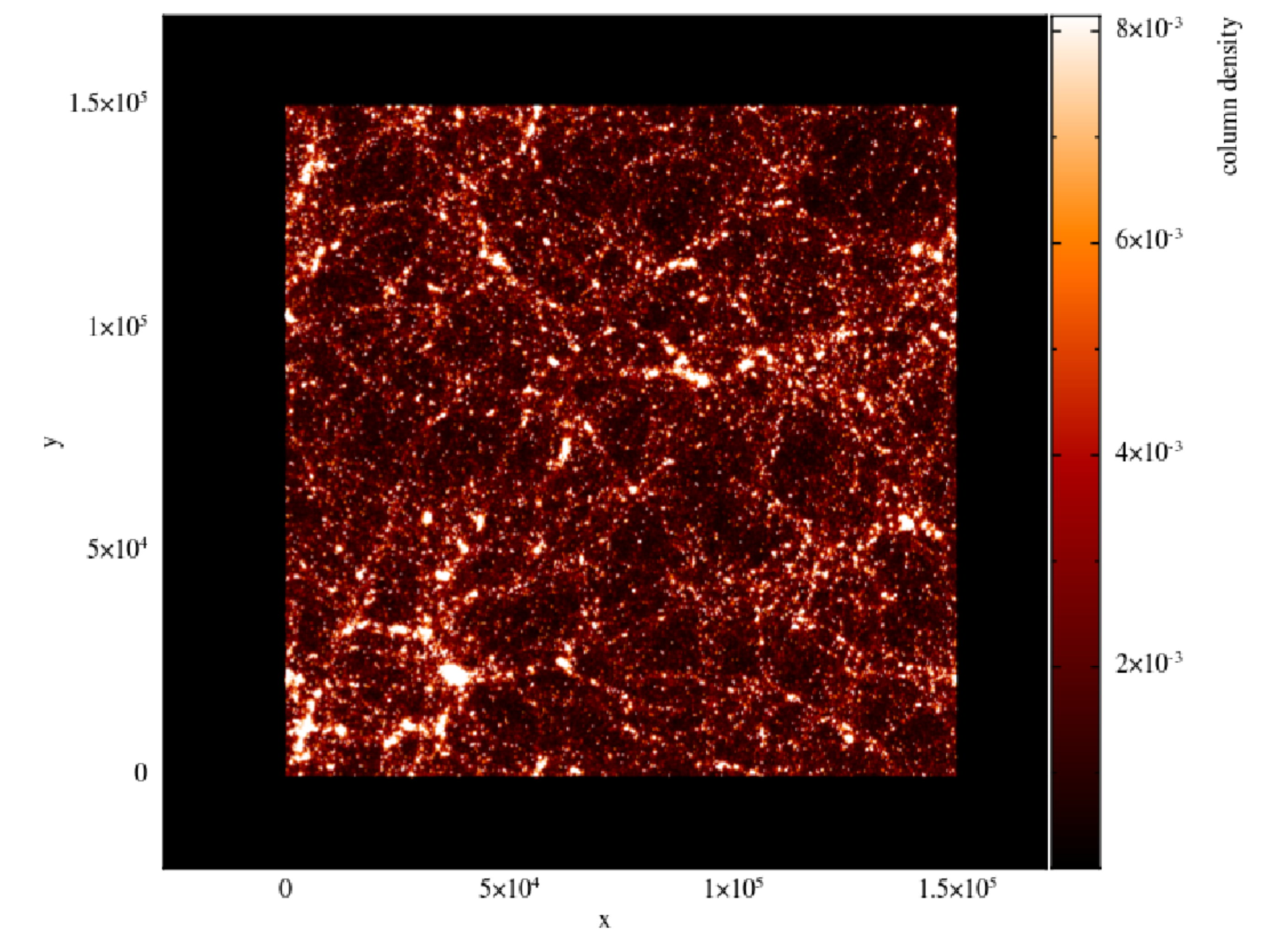}
    \includegraphics[width=0.3\textwidth]{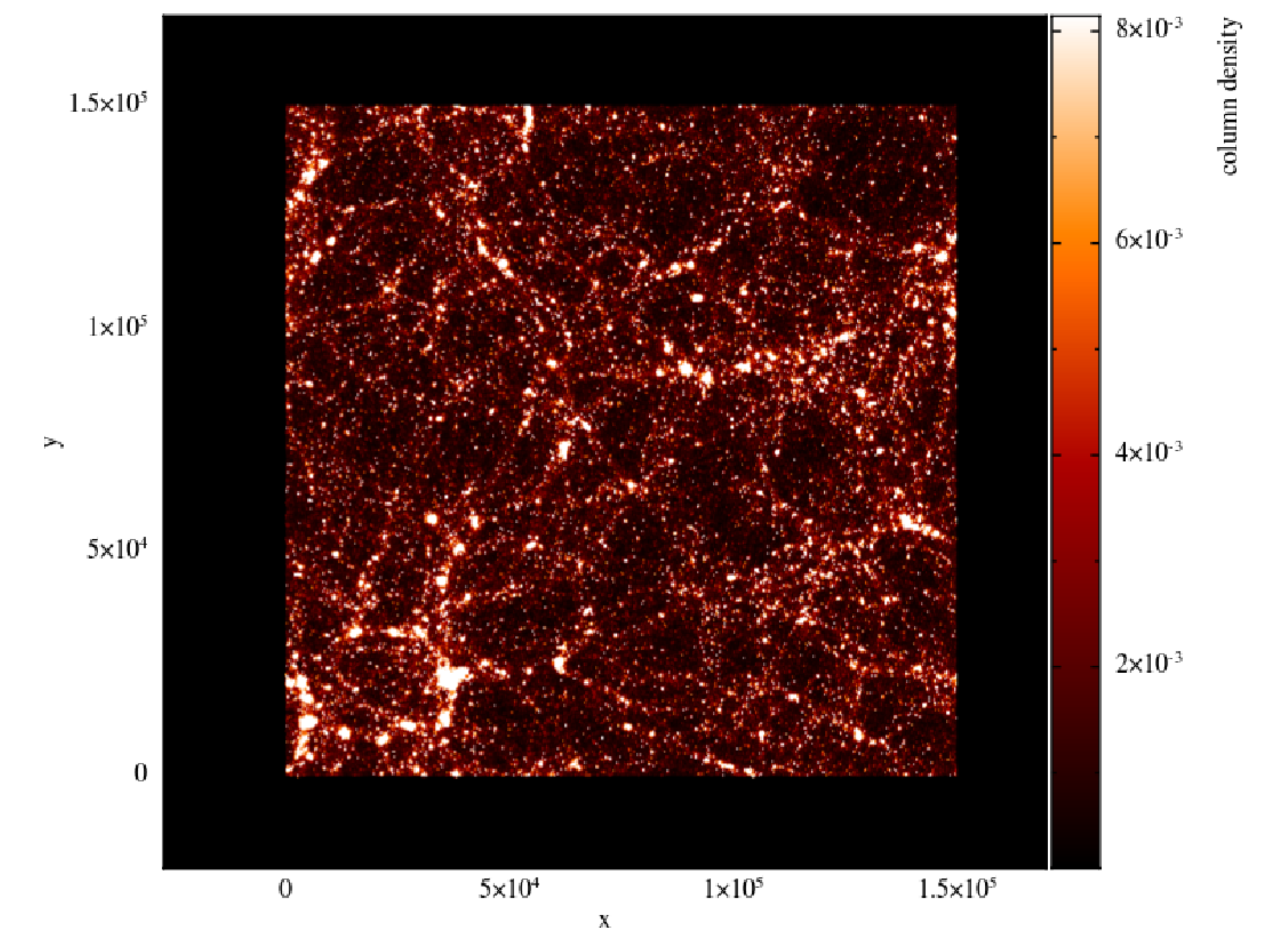}
    \caption{Plano XY de una caja cosmológica de 16777215 partículas desde $z=63$ hasta $z=0$.}
    \label{fig:Simul2}
\end{figure}  

\begin{figure}[htb]
    \centering
    \includegraphics[width=0.45\textwidth]{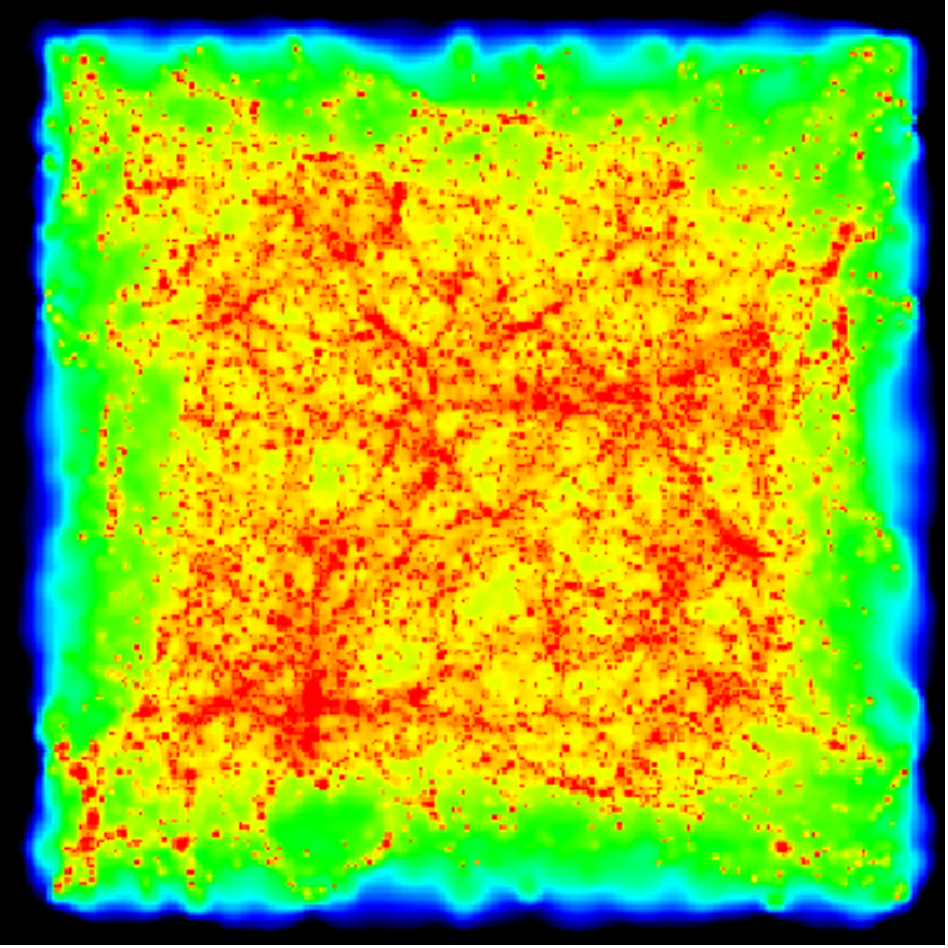}
    \includegraphics[width=0.45\textwidth]{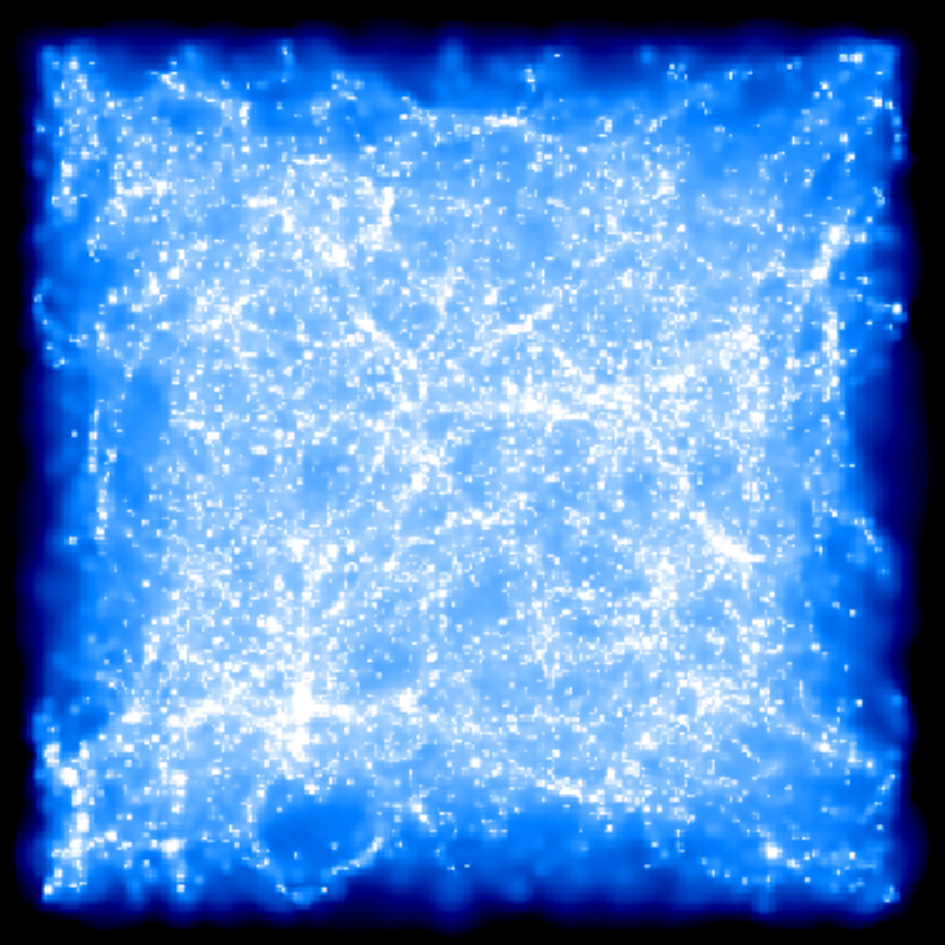}
    \caption{Plano XY del patrón de densidades de la caja cosmológica simulada.}
    \label{fig:Simul3}
\end{figure}  


\newpage

\newpage

\section{Conclusiones}

Como se apreció, la comprensión de la cosmología constituye un reto actual desde diferentes puntos de vista, siendo un campo muy activo observacional, teórica y computacionalmente. Como consecuencia de ello surge un desafío en el que se hace necesario abarcar múltiples herramientas con el fin de modelar y discernir los fenómenos físicos asociados a la compresión misma del universo.

Se obtuvieron las ecuaciones de campo perturbadas en el formalismo métrico, destacando la importancia de los términos perturbativos en el gauge de Poisson. A partir del elemento de línea de FLRW perturbado y tras un tratamiento a las ecuaciones de campo de Einstein, aparece naturalmente la ecuación de campo de Einstein tipo Poisson, que resulta fundamental en la implementación de los algoritmos del código GADGET-2. A su vez la teoría de perturbaciones cosmológicas es entendida como una herramienta útil para estudiar las inhomogeneidades primordiales en cosmología, dando cuenta de la formación y evolución de la estructura del Universo estableciendo una conexión con la cosmología Newtoniana, que justamente se ven representada en los resultados obtenidos de las simulaciones de cajas cosmológicas, al menos a nivel cualitativo.

Se estudi\'o la arquitectura del código GADGET-2, evidenciando que la materia oscura y sistemas estelares se modelan a través de un gas no colisional descrito principalmente por la ecuación de Poisson y la ecuación de Boltzmann no colisional, en donde la función de masa por unidad de volumen, en un campo de velocidades, está dada por una distribución discreta de partículas representativas suavizadas por un kernel. El medio interestelar y el medio intergaláctico se describen  por medio de las ecuaciones de la hidrodinámica con un tratamiento adicional del método SPH. Es relevante mencionar que para el cálculo de fuerzas se utilizó el denominado algoritmo de árbol en conjunto con un método de grillado del espacio para las fuerzas de largo y corto alcance. El avance temporal en la integración numérica utiliza un esquema ``Kick Drift Kick'', que consiste en transformaciones canónicas sucesivas al Hamiltoniano que describe la dinámica de todas las partículas que conforman el universo, al menos dentro del modelo planteado. Se simularon tres cajas cosmológicas con diferentes parámetros bajo el modelo $\Lambda CDM$ desde $z = 63$ hasta $z = 2$, obteniendo resultados interesantes, como la formación de cúmulos  y la ramificación paulatina del sistema a medida que evoluciona hasta un $z$ cercano a cero, que corresponde a la época actual del universo. Los resultados a grandes rasgos están de acuerdo con las simulaciones hechas por el \emph{Virgo Consortium}, en particular la \emph{Millenium Simulation}, en donde se aprecia la estructura típica de cúmulos centrales con ramificaciones que conectan a otros.\\

\renewcommand{\refname}{Bibliograf\'ia}
\bibliographystyle{harvard}
\bibliography{Jorge}

\end{document}